\documentclass[%
 aip,apl,%
 amsmath,amssymb,
 reprint,%
]{revtex4-1}

\usepackage{graphicx}
\usepackage{dcolumn}
\usepackage{bm}
\usepackage{docs}

\draft 

\begin{document}


\title{Surface plasmons at the interface between graphene and kerr-type nonlinear medium} 



\author{Lei Wang}\author{Wei Cai}
\email{weicai@nankai.edu.cn}
\author{Xinzheng Zhang}
\affiliation{The Key Laboratory of Weak-Light Nonlinear Photonics, Ministry of Education, School of Physics and TEDA Applied Physics School, Nankai University, Tianjin 300457, China}
\author{F.~Javier~Garc\'{\i}a de Abajo}
\affiliation{Instituto de \'Optica - CSIC, Serrano 121, 28006 Madrid, Spain}
\author{Jingjun Xu}
\email{jjxu@nankai.edu.cn}
\affiliation{The Key Laboratory of Weak-Light Nonlinear Photonics, Ministry of Education, School of Physics and TEDA Applied Physics School, Nankai University, Tianjin 300457, China}


\date{\today}

\begin{abstract}
The properties of surface plasmons localized at the interface between graphene and kerr-type nonlinear medium in three dimensions are investigated. Compared with surface plasmons at the surface of metal, with the inevitable nonlinear refractive effect, the confinement of plasmon can be improved to three times than graphene plasmons without nonlinear contribution, but also with almost the same relative propagation length. Moreover, the dispersion relation and propation distance of graphene plasmons can be easily controlled by changing the fermi energy, temperature and relaxation time of graphene. Our results suggest a simple but useful potential application for precise nonlinear material sensor using graphene plasmons.
\end{abstract}


\maketitle 

\section{Introduction}

Graphene with unique optical and electronic properties \cite{GN07, Geim09} becomes ideal for a number of photonic application.\cite{LYU11, VE11, EBJ11} In particular, graphene plasmons (GPs), \cite{Mikhai, jablan} which is the collective charge density wave coupled to electromagnetic field localized between graphene, its cover and substrate, have become one of the research hotspots.\cite{JGH11} Compared with surface plasmons in metal, GPs provide a conveniently tunable platform for strong light-matter interactions, much due to their enormous confinement and relatively long propagation length.\cite{abajo1, vidal1} Meanwhile, the induced nonlinear contribution from the substrate is always ignored for metallic plasmons due to that the nonlinear coefficients are always very small. However, this is different for GPs. The strong electromagnetic field of GPs (the enhancement factor is as large as $10^5$-$10^6$) makes the nonlinear contribution should be included when analyzing the properties of graphene plasmons.

As a truly two-dimensional (2D) electronic system, graphene is a monolayer of carbon atoms closed packed in a honeycomb crystal. Previous work verified that graphene system can bind surface plasmons with transverse-magnetic (TM) polarization when Im$\{{\sigma}\}>0$, where $\sigma$ is the conductivity of graphene. \cite{Mikhai} From the fresnel coefficient of the graphene system, the dispersion relation of graphene system is given as:\cite{abajo1,jablan}

\begin{eqnarray}
-i\frac{4\pi\sigma}{c}=\frac{\epsilon}{\sqrt{\eta^{2}-\epsilon}}+\frac{1}{\sqrt{\eta^{2}-1}} \label{eqn1}
\end{eqnarray}
where $c$ is the speed of light in vacuum, $\epsilon$ is the dielectric function of substrate and $\eta$ is the plasmon wave vector normalized to the free light momentum $\omega/c$. Because of the momentum $\eta$ is usually much larger than unit and dielectric function of substrate, so we use the electrostatic limit of this expression and the dispersion of the GPs is obtained

\begin{eqnarray}
\eta\approx\frac{ic(\epsilon+1)}{4\pi\sigma}.\label{eqn2}
\end{eqnarray}

However, when considering the kerr-type nonlinear effect of the substrate, Eq.(\ref{eqn2}) is no longer correct with simply changing the dielectric function $\epsilon$ with the intensity-dependent dielectric expression
\begin{eqnarray}
\epsilon_{d}= \epsilon_0+\alpha \left | \mathbf{E} \right |^2 ,\label{eqn3}
\end{eqnarray}
where $\alpha$ is kerr coefficient of the substrate, $\epsilon_d$ and $\epsilon_0$ are the linear and total dielectric function of the substrate, $\textbf{E}$ is the electric field, respectively. To solve the problem,  in this paper we reconsider the dispersion relation of GPs under the simple system of an infinite planar graphene sheet lying on a dielectric medium, but include the Kerr effect of the substrate, and the analytic express of the dispersion relation of GPs is obtained. Moreover, the dependence of the dispersion relation and propagation distance of GPs on the chemical potential, temperature and relaxation time of graphehe is also clarified.

\section{System description and theory method}

We consider an infinite-large graphene sheet lying on a substrate [Fig. \ref{fig1}(a)], the dispersion property of the GPs at the surface between graphene sheet and substrate can be calculated as follows. Firstly, graphene is physically represented by its in-plane complex conductivity $\sigma$, which can be calculated based on the local random-phase-approximation. The contributions from intraband and interband can be separately expressed as:\cite{vidal1}
\begin{eqnarray}
\begin{aligned}
\sigma_{\text{intra}}&=\frac{2ie^{2}t}{\hbar\pi (\Omega+i\hbar\tau^{-1}/\mu)}\ln\ [2\cosh(\frac{1}{2t})],\\
\sigma_{\text{inter}}&=\frac{e^2}{4\hbar} [\frac{1}{2}+\frac{1}{\pi}\arctan(\frac{\Omega-2}{2t})-\frac{i}{2\pi}\ln \frac{(\Omega+2)^2}{(\Omega-2)^2+(2t)^2}],\nonumber
\end{aligned}%
\end{eqnarray}
where $\Omega=\hbar\omega/\mu$, $t=T/\mu$, with $\mu$ and $\tau$ are the chemical potential and a finite relaxation time of graphene, respectively. Moveover, $T$ is expressed in units of energy. And the total conductivity $\sigma$ can be expressed as
\begin{eqnarray}
\sigma(\omega)=\sigma_{\text{intra}}+\sigma_{\text{inter}}. \label{eqn4}
\end{eqnarray}

Next, for simply the nonlinear response of the substrate is believed to be isotropic, then the electric field in this system can be taken the form as $\mathbf{E}(\mathbf{r},t)=\mathbf{E}(z)\exp(i\beta x-i\omega t)$, with $\beta$ is the momentum of GPs. Substitute this form to Maxwell equations, we can obtain:
\begin{eqnarray}
\left\{\begin{matrix}
iH_{y}=\frac{\partial E_{x}}{\partial \xi}-i\eta E_{z}\\
H_{y}'=i\epsilon_{d} E_{x}\\
-\eta H_{y}=\epsilon_{d} E_{z},
\end{matrix}\right. \label{eqn5}
\end{eqnarray}
where $\xi =kz,\eta =\beta /k$ are normalized length and momentum to the vacuum wave-vector $k$, respectively. The prime is differential to $\xi$. By eliminating the electric field terms in this equation array we can easily get the magnetic equation:
\begin{eqnarray}
(\frac{H_{y}'}{\epsilon_{d}})'=(\frac{\eta^2}{\epsilon_d}-1)H_{y}, \label{eqn6}
\end{eqnarray}
then substitute the electric fields from Eq.(\ref{eqn5}) to Eq.(\ref{eqn3}), one obtains that
\begin{eqnarray}
(\frac{H_{y}'}{\epsilon_{d}})^2=\frac{\epsilon _{d}-\epsilon _0}{\alpha}-(\frac{\eta H_{y}}{\epsilon _{d}})^2, \label{eqn7}
\end{eqnarray}
next differentiate Eq.(\ref{eqn7}) and substitute into Eq.(\ref{eqn6}), then multiply $\epsilon_{d}$ in both sides, one can get the expression:
\begin{eqnarray}
(H_{y}^{2}\ \frac{2\eta^2-\epsilon _d}{\alpha })'=\epsilon_{d}(\frac{\epsilon _{d}-\epsilon _0}{\alpha})',
\end{eqnarray}
integrate over $\xi$ on both sides, we reach the expression:
\begin{eqnarray}
H_{y}^{2}=\frac{\epsilon _d}{2\eta^2-\epsilon _d}\,\frac{\epsilon_d^2-\epsilon_0^2}{2\alpha},\label{eqn9}
\end{eqnarray}
with the equations, the magnetic field can be expressed by dielectric function and normalized momentum.

Assuming that the graphene lies at $z=0$, and the nonlinear medium occupies the region $z>0$. In the nonlinear substrate region the electromagnetic field can be written as
\begin{eqnarray}
\left\{\begin{matrix}
H_{y}=A\\
E_{x}=\frac{H_{y}'}{i\epsilon_{d}}=\frac{A'}{i\epsilon_{d}}\\
E_{z}=-\frac{\eta H_{y}}{\epsilon_{d}}=-\frac{\eta A}{\epsilon_{d}},
\end{matrix}\right.
\end{eqnarray}
also the electromagnetic field in the air layer $z<0$ can be written as
\begin{eqnarray}
\left\{\begin{matrix}
H_{y}=B e^{\kappa \xi}\\E_{x}=\frac{H_{y}'}{i\epsilon}=\frac{\kappa B}{i\epsilon} e^{\kappa \xi}\\
E_{z}=-\frac{\eta H_{y}}{\epsilon_{d}}=-\frac{\eta B}{\epsilon} e^{\kappa \xi},
\end{matrix}\right.
\end{eqnarray}
where $\kappa=\sqrt{\eta^{2}-\epsilon}$ describes the confinement of the system.
From the boundary continuity conditions of $E_x$ and $H_y$ at $\xi=0$, we can know that:
\begin{eqnarray}
\left\{\begin{matrix}
&\frac{A'(0)}{i\epsilon _{d}}=\frac{\kappa B}{\epsilon}\\
& A(0)=(1-\frac{4\pi\sigma\kappa}{ic\epsilon})B
\end{matrix}\right.\label{eqn12}
\end{eqnarray}
 then substitute this equation into Eq.(\ref{eqn7}) at $\xi=0$ we can learn that:
\begin{eqnarray}
\frac{\epsilon_{d}(0)-\epsilon _0}{\alpha}=(\frac{\kappa A(0)}{\epsilon_{d}})^2(1-\frac{4\pi\sigma(\omega)\kappa}{ic\epsilon})^{-2}\nonumber
\\+(\frac{\eta A(0)}{\epsilon _{d}})^2,\label{eqn13}
\end{eqnarray}
combining  Eq.(\ref{eqn9}), Eq.(\ref{eqn12}) and Eq.(\ref{eqn13}), we can get the dispersion equation of GPs finally:

\begin{eqnarray}
-i\frac{4\pi\sigma(\omega)}{c}=\epsilon_{d}(0)\ [\frac{\epsilon _{d}(0)+\epsilon _{0}}{\eta^2(3\epsilon_{d}(0)-\epsilon_0)-2\epsilon_{d}^2(0)}]^{1/2}\nonumber
\\+\frac{\epsilon}{\kappa},\label{eqn14}
\end{eqnarray}
if we set $\alpha\rightarrow 0$ and $\epsilon=1$ ,which means $\epsilon_{d}(0)\rightarrow \epsilon_0$, the Eq.(\ref{eqn14}) will go back to the Eq.(\ref{eqn1}) as expected.

For the case of the normalized momentum $\eta$ is much larger than unit and the dielectric functions of substrate, the wave-vector of GPs can be finally simplified to the following expression:
\begin{eqnarray}
\eta\approx\kappa\approx\frac{ic}{4\pi\sigma(\omega)}[\epsilon_{d}(0)(\frac{\epsilon_{d}(0)+\epsilon _{0}}{3\epsilon_{d}(0)-\epsilon_0})^{1/2}+\epsilon], \label{eqn15}
\end{eqnarray}
where $\epsilon_{d}(0)= \epsilon_0+\alpha\left|\mathbf{E}_0\right|^2$, and $\sigma(\omega)$ takes the form of Eq.(\ref{eqn4}).

\section{Numerical results and discussion}
\begin{figure}[htb]
\centering\includegraphics[width=7cm]{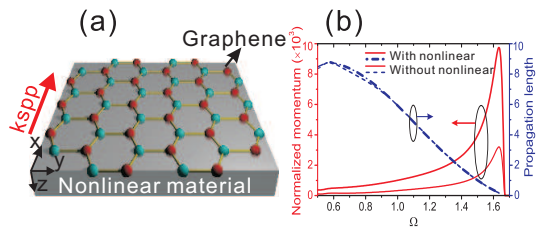}
\caption{(\textbf{a}) Sketch of an infinite planar graphene sheet lying on a substrate. (\textbf{b}) Dispersion relations (solid lines) and propagation distances (dashed lines) of graphene plasmons with (thick lines) and without (thin lines) nonlinear contribution from the substrate. The photon energy is $\Omega=\hbar\omega/\mu$. The momentum is eigenvector in units of $\omega/c$ and the propagation length given by Im($1/{k_{\text{sp}}}$) in unit of SPP wavelength. These magnitudes are obtained for the case of graphene in $\mu=1.0 eV, \tau=1.0\times 10^{-13}s$ and $T=300K$. The nonlinear coefficient is chosen as $1.51\times 10^{-19} m^2/V^2$, and the peak of electric field is set to $E_0^2=1.6\times 10^{20} V^2/m^2$.}
\label{fig1}
\end{figure}

To understand the nonlinear contribution to GPs more clearly, we firstly take the substrate as SiC as an example [Fig.\ \ref{fig1}(a)]. The parameters of graphene and SiC are chosen as follows: the chemical potential, temperature and electron relaxation time are set as 1.0eV, 300K and $1.0\times 10^{-13} s$ for graphene, respectively. The experimental linear dielectric data of SiC is taken from Palik \cite{palik}, while nonlinear coefficient is chosen as $1.51\times 10^{-19}m^2/V^2$ from Weber \cite{Weber} and considered to be non-dispersion. Due to the field enhancement factor is the order of $10^6$ in normal GPs,\cite{abajo1} the peak of electric field can reach $E_0^2=1.6\times 10^{20}V^2/m^2$ at the graphene interface. The dispersion relation and propagation distance of GPs without and with nonlinear contribution from the substrate can be calculated by Eq.\ (\ref{eqn2}) and Eq.\ (\ref{eqn15}) with these parameters, respectively. And the results are shown in Fig.\ \ref{fig1}(b). We can see that the peak momentum of GPs almost can reach 3000 times to free light momentum (thin solid line) when the photon energy (in unit of fermi energy) is 1.6 eV, and the relative propagation length can reach 10 times to SPP wavelength (thin dashed line) without nonlinear contribution of substrate. However, if the nonlinear contribution is taken account, the momentum of GPs reach $10^4$ times to free light wavelength (thick solid line), which means that the nonlinear effect can improve the localization of electromagnetic filed to 3 times. The relative propagation distance of GPs with nonlinear contribution is slightly larger than GPs without nonlinear contribution (thick dashed line), which means that the nonlinear effect has little effect in relative propagation length.

\begin{figure}[htb]
\centering\includegraphics[width=7cm]{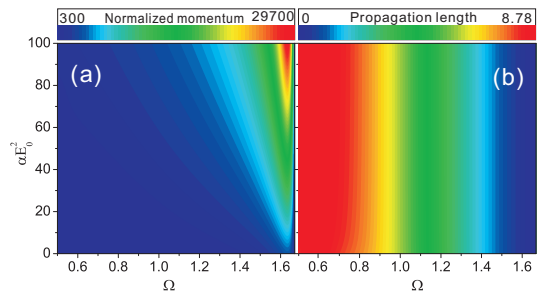}
\caption{The dependence of dispersion relation (\textbf{a}) and relative propagation length (\textbf{b}) of graphene plasmons on the nonlinear dielectric function ($\alpha E^2$) in the system Fig.\ \ref{fig1}(a).  The parameters of graphene are set as $\mu=1.0 eV, \tau=1.0\times 10^{-13}s$ and $T=300K$.}
\label{fig2}
\end{figure}

Moreover, From the Eq.\ (\ref{eqn3}), one can see that the nonlinear part of the dielectric constant of the substrate dependent not only on the nonlinear coefficient, but also on the incident light intensity.  So it is natural to consider how the nonlinear dielectric function ($\alpha E_0^2$) affects the properties of the GPs. The nonlinear dielectric function resolved dispersion and propagation distance of GPs are shown in Fig.\ \ref{fig2}(a) and (b), respectively. One can see that the nonlinear dielectric largely enhances the normalized momentum while affects little on the relative propagation length, which means that it is possible for us to enhance localization of GPs by improving the incident electric field or changing to another material with large nonlinear coefficient.

\begin{figure}[htb]
\centering\includegraphics[width=7cm]{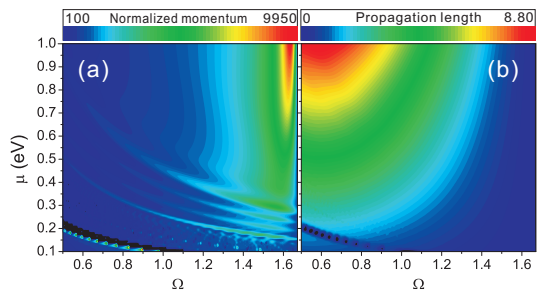}
\caption{ Fermi energy of graphene resolved dispersion relation (\textbf{a}) and relative propagation distance (\textbf{b}) of graphene plasmons with nonlinear contribution from the substrate in the structure shown in Fig.\ \ref{fig1}(a). All of the data are calculated with  $\tau=1\times 10^{-13}s$, $T=300K$, and the nonlinear dielectric function is chosen as $\alpha E_0^2=24.1608$.}
\label{fig3}
\end{figure}

\begin{figure}[htb]
\centering\includegraphics[width=7cm]{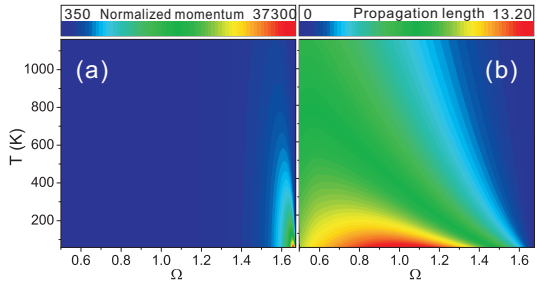}
\caption{ The dispersion relation (\textbf{a}) and relative propagation distance (\textbf{b}) of graphene plasmons in the structure sketched in FIG.\ \ref{fig1}(a) with different environmental temperatures. The parameters of graphene are set as $\mu=1.0 eV$, $\tau=1\times 10^{-13}s$, and the nonlinear dielectric function is $\alpha E_0^2=24.1608$.}
\label{fig4}
\end{figure}

\begin{figure}[htb]
\centering\includegraphics[width=7cm]{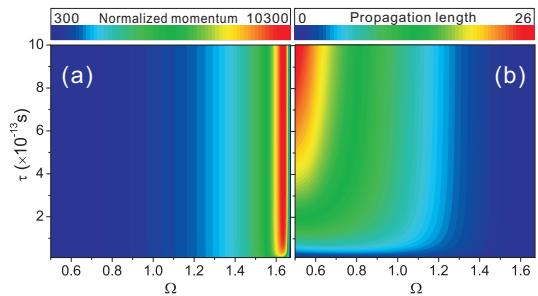}
\caption{ Relaxation time of graphene resolved dispersion relation (\textbf{a}) and relative propagation length (\textbf{b}) of graphene plasmons with the nonlinear contribution from the substrate (SiC) in the structure of Fig.\ \ref{fig1}(a).  These magnitudes are obtained for the case of graphene in $\mu=1 eV$ and $T=300K$, and the nonlinear dielectric function of SiC is set as $\alpha E_0^2=24.1608$.}
\label{fig5}
\end{figure}

In further, compared with metal, the properties of graphene can be controlled by lots of means, such as through doping, applying a gate voltage or changing the temperature. All of these methods can be efficiently change the conductivity of graphene according to Eq.\ (\ref{eqn4}), which open the active regulation of properties of GPs for designing applications. Therefore, three most important physical parameters for describing graphene, chemical potential $\mu$, temperature $T$ and relaxation time $\tau$ are separately studies to look how they affect the properties of GPs with nonlinear contributions in Fig.\ \ref{fig3}, Fig.\ \ref{fig4} and Fig.\ \ref{fig5}, respectively.

Firstly, the fermi energy resolved dispersion relation and propagation length of GPs are shown in Fig.\ \ref{fig3}(a) and Fig.\ \ref{fig3}(b), respectively. The normalized momentum of GPs reaches maximum when the energy approaches 1.6 eV. And the higher the fermi energy is, the larger the normalized momentum and relative propagation are. In addition, the propagation distance of GPs in infrared region is much longer than GPs in the visible region, though they are not so confinement. These results show that higher fermi energy is better for GPs mode, and the application region is more likely lying in the infrared or terahertz region.

Secondly, the environmental temperature can also affect the conductivity of graphene. The collision of electrons in graphene is directly related to the temperature. We calculate the temperature resolved dispersion relation [Fig.\ \ref{fig4}(a)] and propagation length [Fig.\ \ref{fig4}(b)] of GPs. From the contour plots, we know that low temperature benefits to the GPs modes.  In lower temperature, GPs possess more longer propagation distance, which can explained by that the energy loss from collisons of electrons in graphene is reduced. More important, it should be mentioned that the maximum of momentum reaches $3.73\times 10^4$ when the temperature nearly 50 K, this means that GPs with not only comparable long propagation distance but also extremely field enhancement and localization can realize with low temperature in graphene, which can find applications in strong fields physics.

Thirdly, the electron relaxation time $\tau$ is also important for the property of graphene. It is proportion to the fermi energy in ideal condition, but it is not exact correct in the real graphene. Then it opens another freedom to adjust $\tau$ to change the property of GPs. As shown in Fig.\ \ref{fig5}(b), the relaxation time affects the propagation length of GPs which is similar to electrons in metal. The longer the relaxation time is, the longer the plasmon propagates. Fig.\ \ref{fig5}(a) tells us that the relaxation time does not contribute  as much to the dispersion as propagation length of the graphene SPs. Except for that, the relaxation time affects more in the low energy zone than the higher energy region.

\section{Discussion and Conclusion}

Usually, the nonlinear contribution of substrate to GPs is ignored due to the small nonlinear coefficient of materials. However, things have changed for graphene. The very large field localization of GPs leads to super-strong field enhancement. In this case the nonlinear contribution of the material becomes an inevitable factor to calculate the surface plasmon dispersion relation even if the nonlinear coefficient is very small. Alternatively, this property makes GPs potential candidate for the application of precise nonlinear material sensor. In addition, the control of the fermi energy, temperature, electron relaxation time of graphene also make the application of GPs is more easier for sensor.

To conclude, in this paper, the dispersion relation and propagation distance of GPs with nonlinear contribution from the substrate are exactly examined. And compared with the case without nonlinear contribution, we found that the confinement of GPs can be improved to three times with almost the same relative propagation distance. At the same time, the fermi energy, temperature and the relaxation time of graphene also affect the dispersion and relative propagation length. One can find that graphene with high fermi energy, low temperature and low realtime is the best choice for GPs modes, and these modes can propagate longer and have more confinement. The super-high field enhancement can be used in nanoparticle sensor and fluorescence enhancement.



%
%

%

\begin{acknowledgments}
This work is supported by the Fundamental Research Funds for the Central Universities, the National Natural Science Foundation of China (11004112), the National Basic Research Program of China (2007CB307002, 2010CB934101), the 111 Project (B07013), and the Spanish MICINN (MAT2007-66050 and Consolider NanoLight.es).
\end{acknowledgments}


\end{document}